\documentclass[journal=nalefd,email=true,manuscript=letter]{achemso}
\setkeys{acs}{maxauthors = 0}  
\usepackage{times}
\usepackage{graphicx}
\usepackage{siunitx}
\usepackage[version=4]{mhchem} 
\usepackage[T1]{fontenc}       
\usepackage[colorlinks=true,urlcolor=black,linkcolor=black,citecolor=black]{hyperref}
\usepackage{amsmath,amsfonts,amssymb}
\usepackage{soul}

\usepackage{dcolumn}
\usepackage{bm}
\usepackage{tabularx}
\usepackage{latexsym}
\usepackage{changes}
\usepackage{esvect}

\usepackage{mathtools}
\usepackage{xcolor}
\usepackage{soul}
\usepackage{bm}
\usepackage{color}
\usepackage{braket}
\usepackage{dsfont}
\usepackage{wasysym}
\usepackage{xfrac}
\usepackage[utf8]{inputenc}
\usepackage[T1]{fontenc}
\usepackage{mathptmx}

\title{Superconducting contacts to a monolayer semiconductor}

\author{M. Ramezani}
\affiliation{Department of Physics, University of Basel, CH-4056, Basel, Switzerland\\}
\alsoaffiliation{Swiss Nanoscience Institute, University of Basel, CH-4056, Basel, Switzerland\\}
\author{I.\,Correa Sampaio}
\affiliation{Department of Physics, University of Basel, CH-4056, Basel, Switzerland\\}
\author{K. Watanabe}
\affiliation{Research Center for Functional Materials, National Institute for Material Science, 1-1 Namiki, Tsukuba, 305-0044, Japan\\}
\author{T. Taniguchi}
\affiliation{International Center for Materials Nanoarchitectonics,
National Institute for Materials Science, 1-1 Namiki, Tsukuba 305-0044, Japan\\}
\author{C. Sch\"onenberger}
 \affiliation{Department of Physics, University of Basel, CH-4056, Basel, Switzerland\\}
\alsoaffiliation{Swiss Nanoscience Institute, University of Basel, CH-4056, Basel, Switzerland\\}
\author{A. Baumgartner}
\email{andreas.baumgartner@unibas.ch}
 \affiliation{Department of Physics, University of Basel, CH-4056, Basel, Switzerland\\}
\alsoaffiliation{Swiss Nanoscience Institute, University of Basel, CH-4056, Basel, Switzerland\\}

\begin{document}

\begin{abstract}
We demonstrate superconducting vertical interconnect access (VIA) contacts to a monolayer of molybdenum disulfide (MoS$_2$), a layered semiconductor with highly relevant electronic and optical properties. As a contact material we use MoRe, a superconductor with a high critical magnetic field and high critical temperature. The electron transport is mostly dominated by a single superconductor/normal conductor junction with a clear superconductor gap. In addition, we find MoS$_2$ regions that are strongly coupled to the superconductor, resulting in resonant Andreev tunneling and junction dependent gap characteristics, suggesting a superconducting proximity effect. Magnetoresistance measurements show that the bandstructure and the high intrinsic carrier mobility remain intact in the bulk of the MoS$_2$. This type of VIA contact is applicable to a large variety of layered materials and superconducting contacts, opening up a path to monolayer semiconductors as a platform for superconducting hybrid devices.
\end{abstract}

keywords: \emph{TMDC, van der Waals heterostructure, MoS$_2$, monolayer semiconductor, superconducting contacts, superconducting proximity effect}
\newpage

\maketitle


Semiconductors combined with superconducting metals have become a most fruitful field for applications and fundamental research, from gate tunable superconducting qubits \cite{marcuslarsen2015semiconductor}, thermoelectrics \cite{pekolaleivo1996efficient,roddaro2011hot}, to prospective Majorana bound states, \cite{mourik2012signatures,deng2016majorana} or sources of entangled electron pairs \cite{hofstetter2009cooper,hofstetter2011finite, fulop2014local}. These experiments were mainly developed based on one-dimensional (1D) nanowires. To obtain more flexible platforms and scalable architectures, recent efforts focused on two-dimensional (2D) semiconductors \cite{wan2015induced, casparis2018superconducting, fornieri2019evidence,vischi2020electron,graziano2020transport}. However, the number of materials suitable for superconducting hybrids is rather limited. Potentially ideal and ultimately thin semiconductors with a large variety of properties can be found among transition metal dichalcogenides (TMDCs) grown in stacked atomically thin layers. TMDCs often exhibit a broad variety of interesting optical and electronic properties \cite{xiao2012coupled,bhowal2020intrinsic,vzutic2004spintronics}, for example the valley degree of freedom, potentially useful as qubits \cite{wang2012electronics,kormanyos2014spin}, strong electron-electron \cite{qiu2013optical,kormanyos2015k} and spin-orbit interactions \cite{kormanyos2015k}, or crystals with topologically non-trivial bandstructures \cite{tang2017quantum,miserev2019exchange}. One promising material is the semiconductor $\mathrm{MoS}_2$, with a relatively high mobility and large mean free path, allowing for gate-defined nanostructures \cite{pisoni2018interactions,pisoni2018gate, marega2020logic}, which would make MoS$_2$ an ideal platform to combine with superconducting elements.

MoS$_2$ was used as tunnel barrier between superconductors in vertical heterostructures \cite{island2016thickness,trainer2020proximity} and showed signs of intrinsic superconductivity \cite{costanzo2016gate} and of a ferromagnetic phase at low electron densities \cite{roch2019}. However, to exploit the intrinsic properties of MoS$_2$ and to fabricate in-situ gate tunable superconducting hybrid structures, direct superconducting contacts in lateral devices are required. Such contacts are difficult to fabricate due to the formation of Schottky barriers \cite{cui2015multi,allain2015electrical,tang2017quantum}, material degradation \cite{Schauble_Pop_ACSNano_2020} and fabrication residues when using standard fabrication methods \cite{Samm2014, lembke2015single,pisoni2018gate}. Less conventional edge contacts were also found problematic recently \cite{Seredinski_Finkelstein_arXiv_2021}.

Here, we report vertical access interconnect (VIA) contacts \cite{viatelford2018} to monolayer MoS$_2$ with the superconductor MoRe as contact material. We demonstrate a clear superconducting gap in the transport characteristics, including the magnetic field and temperature dependence, and features suggesting stronger superconductor-semiconductor couplings, forming the basis for superconducting proximity effects and bound states. In addition, we show that this fabrication method retains the intrinsic MoS$_2$ bulk properties, including a large electron mobility and sequentially occupied spin-orbit split bands \cite{pisoni2018interactions}.

\begin{figure}[t]
\includegraphics[width=0.6\columnwidth]{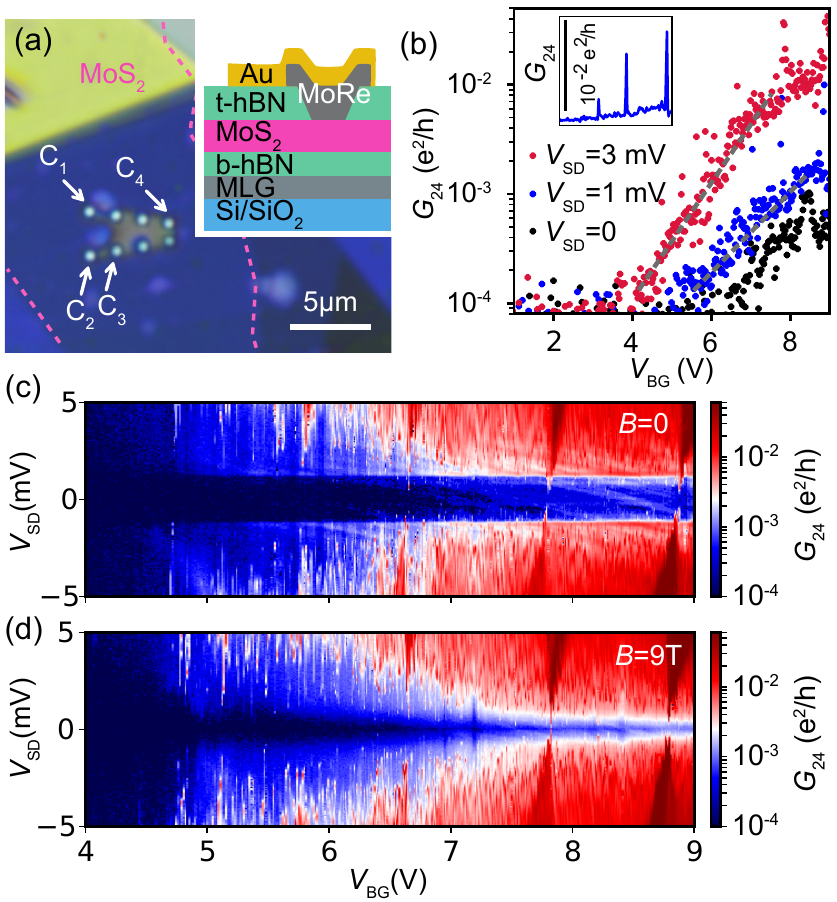}
\caption{\label{fig1} (a) Optical microscopy image of the $\mathrm{MoS}_2$ heterostructure, with MoRe VIA contacts pointed out by white arrows. Inset: schematic of a VIA contact, with t-hBN and b-hBN denoting the top and bottom hBN layers, respectively. (b) Differential conductance $G_{24}$ as a function of $V_{\rm BG}$ for a series of $V_\mathrm{SD}$ on a logarithmic scale, and as an inset for $V_\mathrm{SD}=1\,$mV on a linear scale. (c) $G_{24}$ vs. $V_{\rm BG}$ and $V_{\rm SD}$ at $B=0$ and (d) at $B=9\,$T.}
\end{figure}

Figure~\ref{fig1}(a) shows an optical microscopy image of the presented device, and a schematic of a single VIA contact. The MoS$_2$ is fully encapsulated by exfoliated hexagonal boron nitride (hBN), ensuring minimal contamination of the bulk materials, while the following assembly process allows the fabrication of pristine material interfaces and contacts:\\
\newline
1) vertical access: using electron beam lithography (EBL), VIA areas with a radius of $200\,$nm are defined on the designated top hBN flake ($\sim 40\,$nm thickness) on a Si/SiO$_2$ wafer, and etched completely open by reactive ion etching with a 20:5:5 sccm SF$_6$:O$_2$:Ar mixture at $25\,$mTorr chamber pressure and $50\,$W RF power.\\
2) VIA metalization: in a second EBL step, a slightly larger area with the VIA in the center is defined for mechanical anchoring to the top hBN. We then deposit the type II superconductor MoRe (bulk critical temperature $T_{\rm c}\approx 6-10\,$K, (second) critical magnetic field $B_{\rm c}\approx 8-9\,$T \cite{more1,sundar2013electrical}) using sputter techniques. As the optimal film thickness we find $10\,$nm plus the top hBN thickness.\\
3) Stacking of layers: the wafer with the VIA structure is transferred to an inert gas (nitrogen) glove box, where the top hBN layer with the metalized VIAs is picked up from the substrate using a polycarbonate (PC) stamp and an hBN helper layer, and then used to pick up consecutively a monolayer MoS$_2$ flake, a bottom hBN flake ($\sim 25\,$nm thickness), and a multilayer graphene (MLG) flake serving as backgate. \\
4) Finish: the stack is then deposited onto a Si/SiO$_2$ wafer, where macroscopic Ti/Au ($10/50\,$nm) leads to the VIAs are fabricated using EBL. The sample is then annealed at 350$^\circ$C for 30\,min, in a vacuum chamber with a constant flow of forming gas. \\

Using gold as VIA material, this fabrication process yields $> 80\%$ of the contacts with two-terminal resistance-area products smaller than $200\,$k$\Omega\mu{\rm m}^{2}$ at $T=1.7$K at a backgate voltage of $V_{\rm BG}=10\,$V. This yield is reduced to roughly $50\%$ when using MoRe, possibly due to a material loss during the pick-up procedure. In the presented device, only half of the contacts show resistances lower than $200\,\mathrm{k}\Omega\mu{\rm m}^{2}$, which we use for the experiments and are labeled from C$_1$ to C$_4$ in Fig.\,\ref{fig1}(a). The two terminal differential conductances $G_{jk}=\mathrm{d}I_j/\mathrm{d}V_k$ we obtain by measuring the current variation in the grounded contact C$_j$ while applying a modulated bias voltage $V_\mathrm{SD}$ to contact C$_k$ using standard lock-in techniques. The experiments were performed in a dilution refrigerator at $\sim 60\,$mK, while for higher temperatures, we used a variable temperature insert (VTI) with a base temperature of $\sim 1.7\,$K. In addition, we apply an external magnetic field $B$ perpendicular to the substrate.

In Fig.\,\ref{fig1}(b), the differential conductance $G_{24}$ is plotted as a function of the backgate voltage $V_\mathrm{BG}$ for several bias voltages $V_{\rm SD}$. Increasing $V_\mathrm{BG}$ results in an exponential increase in $G_{24}$, starting from a pinch-off voltage of $V_\mathrm{BG}\approx 6\,$V for $V_{\rm SD}=0$. This value is offset towards smaller values for larger bias voltages, a first indication for an energy gap. However, as seen in the inset of Fig.\,\ref{fig1}(b), in this regime we find very sharp peaks in $G$, consistent with Coulomb blockade (CB) effects. We note that an increase in $V_\mathrm{BG}$ not only changes the charge carrier density in the MoS$_2$, but also the Schottky barrier at the metal-semiconductor interface and disorder induced charge islands.
To demonstrate a superconducting energy gap and to distinguish it from other effects like CB, we plot in Fig.\,\ref{fig1}(c) the conductance $G_{24}$ vs. $V_{\rm SD}$ over a large range of $V_{\rm BG}$ at $B=0$, while Fig.\,\ref{fig1}(d) shows the same experiment at $B=9\,$T. At $B=0$, independent of the gate voltage, one clearly finds a strongly suppressed conductance for roughly $|V_{\rm SD}|< 1.2\,$mV, a gap size consistent with literature values for the superconducting energy gap of MoRe \cite{singh2014molybdenum}. The conductance is suppressed by a factor $\sim 8$ between the large and the zero bias values at $V_{\rm BG}\approx 6\,$V, and by a factor of $\sim 15$ near $V_{\rm BG}= 8\,$V. We note that such a sharp gap is only observable if a tunnel barrier is formed between the semiconducting MoS$_2$ and the superconducting region, at least at one contact. The discrete features inside the gap are probably not Andreev bound states \cite{Pillet_NatPhys_2010, Gramich_PRB_2017}, but rather originate from gate-modulated conductance features in the bulk MoS$_2$. At $|V_{\rm SD}|>1.2\,$mV, we find a strong modulation of $G$, which we interpret as several Coulomb blockaded regions.
These resonances suggest that there is significant disorder near some of these contacts, so that we can think of this device as an MoS$_2$ region, incoherently coupled to the contacts by two normal-superconductor (N/S) junctions. The reason for one junction, namely the less transparent one, dominating the transport characteristics is that the junction with the higher transmission has a  reduced resistance in the energy gap due to Andreev reflection, in which two electrons are transferred to S to form a Cooper pair.
At $B=9\,$T, the superconducting gap is reduced, as discussed below in detail, but we now find that the gap becomes gate dependent. While in short semiconducting nanowires a gate tunable proximity effect was reported \cite{junger2019spectroscopy, heedt2020shadow}, we tentatively attribute our finding to a gate independent superconducting energy gap convoluted with a gate tunable MoS$_2$ conductance.

\begin{figure}[ht!]
\includegraphics[width=0.6\columnwidth]{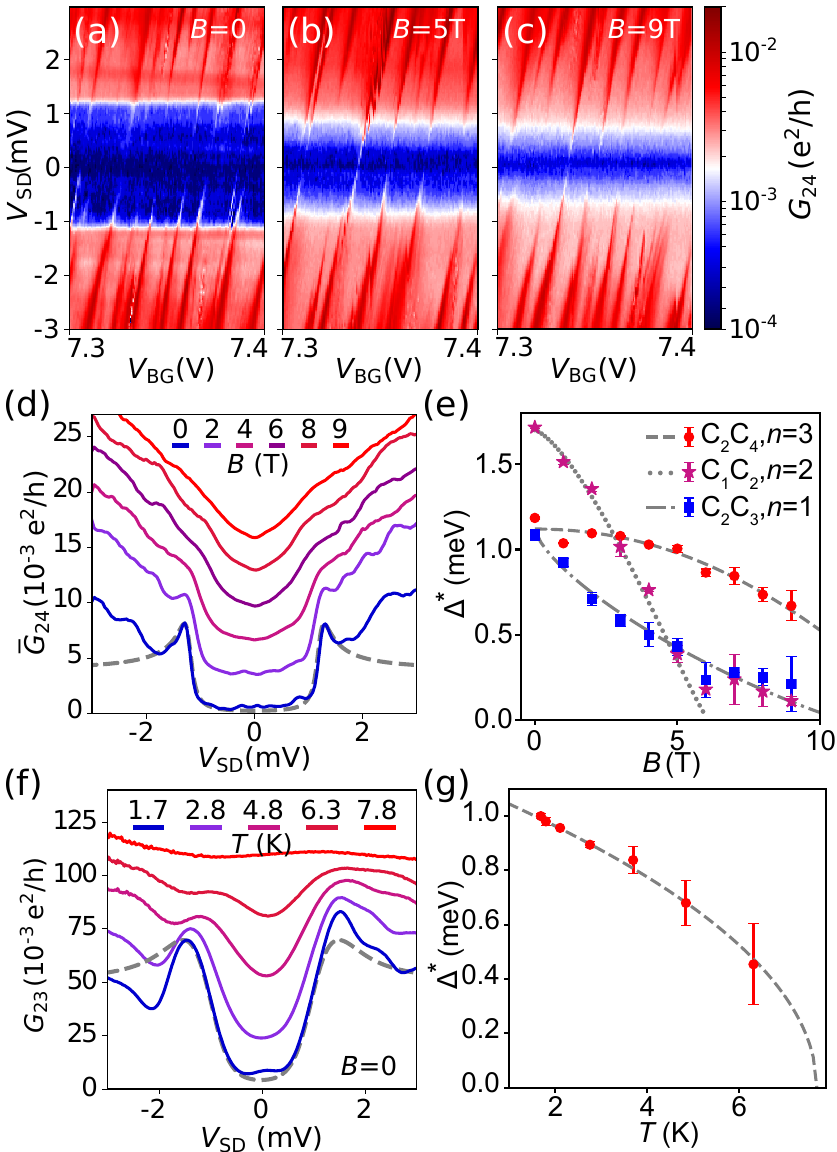}
\caption{\label{fig2} (a) Differential conductance $G_{24}$ as a function of the bias $V_{\rm SD}$ and the backgate voltage $V_{\rm BG}$ at $B=0$, (b) $B=5\,$T, and (c) $B=9\,$T. (d) $G_{24}$ averaged over a $V_{\rm BG}$ interval of $0.5\,$V plotted vs. $V_{\rm SD}$ for the indicated magnetic fields, with the curves offset for clarity. (e) Superconducting energy gap $\Delta^{*}$ as a function of $B$ for different contact pairs. $\Delta^{*}$ is extracted from the inflection points in the curves of Fig.~2(d) [red disks], from Fig.~3 [purple stars] and from an additional data set discussed in the supplemental data, Fig.~S1 [blue rectangles]. (f) $G_{23}$ vs. $V_{\rm SD}$ recorded at the indicated temperatures $T$. (g) $\Delta^{*}$ vs. $T$ extracted from the data in (f). All theoretical curves (dashed and dotted lines) are discussed in the text.}
\end{figure}

We investigate the gap in the transport characteristics and the field dependence in more detail in Fig.~\ref{fig2} for the same contact pair. Figures~\ref{fig2}(a)-(c) show higher resolution conductance maps for a smaller $V_{\rm BG}$ interval, for the magnetic fields $B=0$, $B=5\,$T and $B=9\,$T, respectively. The figures show a very clear gap in the conductance map, which is systematically reduced with increasing magnetic field, independently of the sharp resonances. The positions of the latter are unaffected by the gap, so that we can attribute them to resonances in the MoS$_2$, for example due to CB. To extract the energy gap, we plot $G_{24}$ in Fig.~\ref{fig2}(d), averaged over a gate voltage interval of $0.5\,$V for each $V_{\rm SD}$ value, for a series of perpendicular magnetic fields. These curves show how the energy gap closes with increasing $B$. The curve at $B=0$ can be fitted well using the model by Blonder, Tinkham and Klapwjik (BTK) \cite{blonder1982transition}, including an additional broadening parameter \cite{dynes1978direct}, as shown in Fig.\,\ref{fig2}(d) by a gray dashed line. The fit parameters are consistent with a weakly transmitting barrier in a single N/S junction. At larger fields, the extracted parameters become ambiguous due to a strong broadening of the curves. As a measure for the superconducting energy gap $\Delta^{*}$, we therefore plot in Fig.\,\ref{fig2}(e) the average of the low-bias inflection points of each curve (red dots). For $B=0$, we find $\Delta^{*}_{0}\approx 1.2\,$meV, in good agreement with bulk MoRe \cite{island2016thickness,singh2014molybdenum} and one S/N junction dominating the transport. The field dependence of $\Delta^{*}$ is well described by standard theory of superconductivity for pair breaking impurities in a metal with a mean free path shorter than the superconducting coherence lengths. For the corresponding self-consistency equations we use $\Delta^{*}(\alpha)=\hat{\Delta}(\alpha)[1-(\alpha/\hat{\Delta}(\alpha))^{2/3}]^{3/2}$, with $\Delta^{*}(\alpha)$ the energy gap in the excitation spectrum and $\hat{\Delta}$ the order parameter determined numerically from $\ln (\hat{\Delta}(\alpha)/\Delta_{0})= - \pi\alpha /4\hat{\Delta}(\alpha)$ for a given pair breaking parameter $\alpha$ \cite{Skalski_PR_1964, gramich2016subgap}. The latter we interpolate as $\alpha=0.5 \Delta_{0}^{*}(B/B_\mathrm{c})^{n}$, with $n$ a characteristic exponent. As shown in Fig.\,\ref{fig2}(e), the best fit we obtain for $n=3$, $\Delta_{0}^{*}= 1.12\,$meV and the (upper) critical field $B_{\rm c}=14.5\,$T. The latter value is clearly larger than reported for bulk MoRe. Seemingly similar data plotted as purple stars and blue rectangles we discuss below.

As expected for a superconducting energy gap, $\Delta^{*}$ is also reduced with increasing temperature $T$. In Fig.\,\ref{fig2}(f) we plot $G_{23}$ as a function of $V_{\rm SD}$ at $V_{\rm BG}=9\,$V for a series of temperatures at $B=0$ in a different cooldown. For the lowest values of $T$, we can reproduce the data using very similar BTK and Dynes parameters as above, only adjusting the normal state resistance and the temperature to $T=1.7\,$K. However, at higher temperatures, the fits become ambiguous, due to a broadening and possibly a temperature dependence of the Schottky barrier \cite{cui2015multi}. Again, we plot the inflection points of the curves as an estimate for $\Delta^{*}$, as shown in Fig.\,\ref{fig2}(g). To determine the critical temperature, we fit the expression $\Delta^{*}=\Delta^{*}_0\sqrt{1-T/T_{\rm c}}$ and find  $T_{\rm c}=7.7\,$K and $\Delta^{*}=1.2\,$meV, consistent with literature values for bulk MoRe \cite{singh2014molybdenum,sundar2013electrical}.

\begin{figure}[ht!]
\includegraphics[width=0.6\columnwidth]{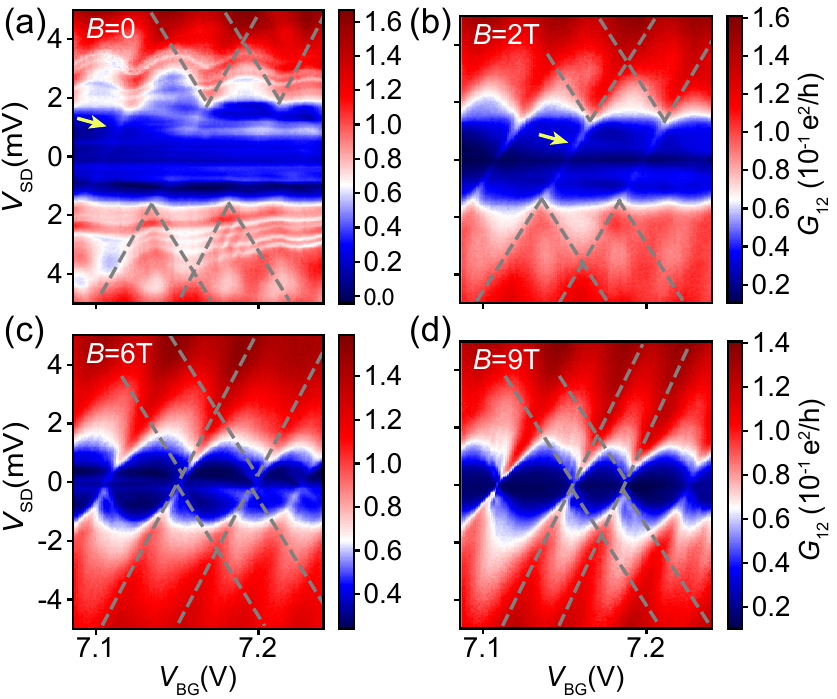}
\caption{\label{fig3} $G_{12}$ plotted as a function of $V_{\rm BG}$ and $V_{\rm SD}$ at (a) $B=0$, (b) $B=2\,$T, (c) $B=6\,$T and (d) $B=9\,$T, recorded at $T=60\,$mK. The dashed lines trace the CB diamonds, shifted vertically and horizontally between the subfigures, while the yellow arrows point out lines of resonant Andreev tunneling.}
\end{figure}

Up to this point, our experiments demonstrate superconducting contacts, but with a weak electronic coupling between the MoS$_2$ and the reservoirs, at least for one contact junction. However, we also find evidence for a stronger coupling of MoS$_2$ regions to the superconductors, relevant for devices relying on the superconducting proximity effect. As an example, we show bias spectroscopy measurements with CB features between contacts C$_1$ and C$_2$ in Fig.~\ref{fig3}, for a series of magnetic fields $B$. Similarly as in Fig.~2, we find a transport gap, reduced by larger $B$ values. Here, the low-bias ends of the CB diamonds are shifted in energy and in gate voltage, as indicated by the gray dashed lines, consistent with a MoS$_2$ quantum dot (QD) directly coupled to one superconducting contact (i.e. forming an S-QD-N junction)  \cite{gramich2015resonant}. These tips of the CB diamonds are connected across the gap by a single faint resonance, pointed out by yellow arrows, best seen in Fig.~3(b) at $B=2\,$T. We attribute these lines to resonant Andreev tunneling \cite{gramich2015resonant}, in which the electrons of a Cooper pair pass through the QD in a higher order tunneling process. This process is suppressed much stronger by a tunnel barrier than single particle tunneling \cite{sun1999resonant}, which suggests that the QD is strongly coupled to the superconductor. At large $B$ fields, a quasi-particle background in the superconducting density of states results in single particle CB diamonds \cite{junger2019spectroscopy}. With a QD charging energy of $\sim 2\,$meV and using $E_\mathrm{c}=\frac{e^{2}}{8\epsilon\epsilon_{0}r}$ for a disc shaped QD encapsulated in hBN, we estimate the radius of the confined QD region as $r\approx 300\,$nm.

The shift of the CB diamonds in $V_{\rm SD}$ gives a measure for $\Delta^{*}$ \cite{gramich2015resonant}, which we read out at the bias at which $\sim 50\%$ of the large bias conductance is reached at the tip of the CB diamond. The extracted values are plotted as purple stars in Fig.~2(e). Surprisingly, we find a significantly larger zero field gap, $\Delta_0^{*}\approx 1.7\,$meV, and a rather different functional dependence on $B$ than for the curves analyzed in Fig.~2 (red dots). The latter is demonstrated by the dotted line obtained for the exponent $n=2$, and $B_{\rm c}=6.4\,$T. In addition, Fig.~2(e) shows a third $\Delta^{*}$ curve extracted from CB diamond shifts in experiments on another contact pair shown in the supplemental data, Fig.~S1. For this curve, we obtain $n=1$, while $\Delta^{*}\approx 1.12\,$meV and $B_{\rm c}\approx 12\,$T correspond well to the previously obtained values.

While a larger gap in the transport experiments can be simply attributed to a significant fraction of the bias developing across another part of the device, for example across the second N/S junction, the different functional dependence is more difficult to explain. Since nominally the geometry and MoRe film thickness are identical for all contacts, we tentatively attribute this finding to a superconducting proximity region forming in the MoS$_2$ near a strongly couped contact, yielding an induced superconducting energy gap $\Delta^{*}$ \cite{junger2019spectroscopy}, and a different $B$-field dependence of the pair breaking compared to the bulk superconductor.

Additional indications for a stronger coupling to S are the almost gate voltage independent features at $B=0$, reminiscent of two NS junctions and multiple Andreev reflection processes in between, with a much stronger suppression with increasing $B$ than for the observed gap. In the supplemental data, Fig.~S2(a), we also show data at higher gate voltages, exhibiting a conductance minimum instead of a maximum at the bias that corresponds to the energy gap, developing into negative differential conductance at the lowest field values. These findings are qualitatively consistent with calculations for an S/I/N/S junction with resonances in the N region \cite{Zhitlukhina_2016}.

\begin{figure}[ht!]
\includegraphics[width=0.6\columnwidth]{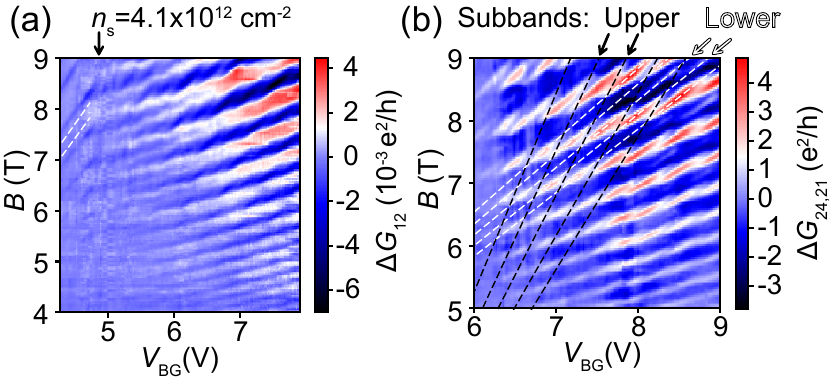}
\caption{\label{fig4} (a) Two-terminal dc conductance $G_{12}=I_{1}/V_{\rm SD}$ with $V_{\rm SD}=8\,$mV  applied to contact C$_2$, plotted as a function of the magnetic field $B$ and the backgate voltage $V_{\rm BG}$ at $T\approx 60\,$mK. $n_{\rm s}$ points out the gate voltage corresponding to the electron density at which the higher spin-orbit subbands start to be populated. (b) Three-terminal dc conductance $G_{24,21}=I_{2}/V_{12}$, with an external bias $V_{\rm SD}=10\,$mV applied to C$_4$, while the current $I_2$ is measured at C$_2$ and the voltage difference $V_{12}$ between C$_1$ and C$_2$. In both maps, a third order polynomial was subtracted at each gate voltage to remove a smooth background.}
\end{figure}

The above data show that our fabrication scheme results in superconducting contacts to monolayer MoS$_2$, possibly with a reasonably strong coupling to the superconductors for some of the contacts. To demonstrate that the intrinsic properties of MoS$_2$ are intact in the bulk crystal, we investigate quantum transport in large magnetic fields and at bias voltages large enough to render the superconducting energy gap irrelevant. In Fig.\,\ref{fig4}(a) we plot the dc conductance $G_{12}=I_{1}/V_{\rm SD}$ as a function of $V_{\rm BG}$ and $B$, at $V_{\rm SD}=8\,$mV and $T=60\,$mK, from which we have subtracted a third order polynomial background for each gate voltage to eliminate effects from the classical Hall effect and CB effects.

Figure\,\ref{fig4}(a) shows well developed Shubnikov de Haas (SdH) oscillations, suggesting a high $\mathrm{MoS}_2$ quality, with an onset at $B_{\rm on} < 4\,$T. In the Drude model, this onset is interpreted as the charge carriers closing a cyclotron orbit before being scatted, which happens roughly at $\omega_{\rm c}\tau>1$, with $\omega_{\rm c}=eB/m^{*}$ the cyclotron frequency, $m^{*}$ the effective electron mass, and $\tau$ the scattering time. This yields a lower bound for the carrier mobility of $\mu=e\tau/m^{*}\approx 2500\,$cm$^2$/(Vs), similar to $\mu\approx 5000\,$cm$^2$/(Vs) we obtain with Au VIA contacts, identical to the best literature values \cite{pisoni2018gate}. The discrepancy in mobility between the MoRe and the Au VIA contacts we attribute to heating effects due to the much larger bias we apply to the MoRe contacts to avoid effects of the superconducting energy gap.

The quality of the MoS$_2$ can also be seen in the fact that the the four lowest spin-orbit subbands, corresponding to the valley and the spin degree of freedom, are not mixed by disorder. We find that the slope of the SdH resonances changes by roughly a factor of two at $V_\mathrm{BG}^{*}=4.8\,$V, corresponding to an electron density of $n_\mathrm{s}\approx 4.1 \times 10^{12} \mathrm{cm}^{-2}$, at which the two upper spin-orbit subbands become populated. Using $m^{*}=0.6$, we obtain a subband spacing of $\sim 15\,$meV, as reported previously \cite{pisoni2018interactions}. These features prevail also at $T\approx 1.7\,$K, as shown in the supplemental data, Fig.~S2(b).

The two-terminal magneto-conductance measurements suffer from large background resistances due to Shottky barriers, which we can partially circumnavigate by performing a three terminal experiment. In Fig.~4(b) we plot the dc conductance $G_{24,21}$, as explained in the figure caption. This technique removes the contact resistance at C4, so that the conductance resonances due to the Landau levels can be measured more clearly. The results in Fig.~4(b) show similar patterns as in better suited Hall bar experiments \cite{pisoni2018interactions}, exhibiting clear superposition patterns of the spin and valley split subbands, indicated by dashed lines. We note that due to the less ideal contact geometry of our devices, we cannot go to lower electron densities in these experiments, because the current density passing near the remote contacts is very low.

In conclusion, we established superconducting contacts to a monolayer of the TMDC semiconductor MoS$_2$ using vertical interconnect access (VIA) contacts, and characterized the superconducting energy gap in different transport regimes. The fact that in most experiments one N/S junction dominates the transport characteristics, and signatures of resonant Andreev tunneling and a superconducting proximity effect, suggest that also contacts with a stronger transmission between the superconductor and the semiconductor are possible, thus opening a path towards semiconductor-superconductor hybrid devices at the limit of miniaturization, with a group of materials - the TMDCs - that offers a very large variety of material properties and physical phenomena.

\section*{Associated content}
The supplemental data mentioned in the text as Figs.~S1 and S2 are available free of charge on the ACS Publications website at DOI:\\

All data in this publication are available in numerical form at
\href{http://doi.org/10.5281/zenodo.4518683}{doi.org/10.5281/zenodo.4518683}

\section*{Acknowledgments}
This work was supported financially by the Swiss National Science Foundation project "Two-dimensional semiconductors for superconductor hybrid nanostructures" granted to AB, the Swiss Nanoscience Institute (SNI) project P1701, and  the   ERC   project   Top-Supra  (787414).   K.W.  and  T.T.  acknowledge  support  from  the  Elemental  Strategy  Initiative  conducted  by  the  MEXT,  Japan,  Grant  Number JPMXP0112101001,  JSPS  KAKENHI  Grant  Numbers JP20H00354 and the CREST (JPMJCR15F3), JST.

\bibliography{references}

\end{document}